\documentclass{article}
\usepackage{amsmath,amssymb,graphicx}

\newcommand{\ket}[1]{| #1 \rangle}
\newcommand{\braket}[2]{\langle #1 | #2 \rangle}

\title{Entropic uncertainty relations for incomplete sets of mutually unbiased observables}
\author{Adam Azarchs, California Institute of Technology}

\begin{document}
\maketitle

\begin{abstract}
Entropic uncertainty relations, based on sums of entropies of probability distributions arising from
different measurements on a given pure state, can be seen as a generalization of the Heisenberg
uncertainty relation that is in many cases a more useful way to quantify incompatibility between
observables.  Of particular interest are relationships between `mutually unbiased' observables, which
are maximally incompatible.  Lower bounds on the sum of entropies for sets of two such observables,
and for complete sets of observables within a space of given dimension, have been found.  This paper
explores relations in the intermediate regime of large, but far from complete, sets of unbiased
observables.
\end{abstract}

\section{Background and prior work}

Entropic uncertainty relations are analogous to the Heisenberg uncertainty principle, but phrased in terms of the
entropy of incompatible measurements rather than the variances.  The entropic form is preferable to the original
Heisenberg formulation in many situations.  The entropy of measuring a state $\ket{\psi}$ with respect to
basis $\{\ket{a_i^{(k)}}\}_{i=1}^N$ is defined for these purposes as
$$ H_k = -\sum_{i=1}^N p_i^{(k)} \log p_i^{(k)}, $$
where $N$ is the dimension and $p_i^{(k)}$ is the probability of the measurement yielding outcome $i$.
That is, 
$$ p_i^{(k)} = | \braket{a_i^{(k)}}{\psi} |^2. $$

David Deutsch \cite{deutsch} found a relationship
$$H_a + H_b \geq - 2 \log \frac{1}{2} (1+c),$$
where
$$c = \max_{j,k} |\braket{a_j}{b_k}|.$$
Maassen and Uffink \cite{maassen} improved this bound to
\begin{equation}
\label{eq:pair}
H_a + H_b \geq - 2 \log c.
\end{equation}
This later relationship is much stronger since $c \ll 1$ for large $N$.

Of particular interest are so-called mutually unbiased observables (or bases), defined as those where
$$ |\braket{a_i^{(k)}}{a_j^{(l)}}|^2 = 1/N $$
for all $k \neq l$.  It has been shown \cite{ivonovic} that at least for prime $N$ there exist sets of $N+1$
such observables.
S\'{a}nchez \cite{sanchez} found that for such sets, one could write
\begin{equation}
\label{eq:san}
\sum_{k=1}^{N+1} H_k \geq (N+1) \log \frac{1}{2}(N+1).
\end{equation}
This is substantially better than the bound derived by breaking up the set pair-wise and
using equation \ref{eq:pair},
\begin{equation}
\label{eq:weakfull1}
\sum_{k=1}^{N+1} H_k \geq \frac{1}{2} (N+1)\log N.
\end{equation}

For several interesting applications, such as encryption, a relationship like (\ref{eq:san})
summing only over the incomplete set of $M$ observables, would be useful.  Weak lower bounds can be formed
by modification of (\ref{eq:weakfull1}) to
\begin{equation}
\label{eq:weak1}
\sum_{k=1}^M H_k \geq \frac{1}{2} M \log N.
\end{equation}
However, this quickly becomes a poor bound as $M$ increases.
Similarly, for $M$ approaching $N+1$ we can take (\ref{eq:san}) and subtract the upper bound on the entropy of
the observables that are being left out, $(N+1-M) \log N$, which yields
\begin{equation}
\label{eq:weak2}
\sum_{k=1}^M H_k \geq (N+1) \log \frac{N+1}{2N} + M \log N.
\end{equation}
However for $M \ll N$ this is actually negative, so useless in that regime.
The regime where $N$ is very large, and $M$ is also large but $\ll N$, is quite interesting to `locking' correlations 
\cite{locking1}.  A relation useful in the domain around $M \sim N/\log N$ is derived in this paper.

\section{Result for intermediate domain}
Larsen \cite{larsen} has derived the equality relationship
$$\sum_{k=1}^{N+1} \pi_k = \Pi + 1$$
where the $\pi_k$s are the `purities,'
$$\pi_k = \sum_{i=1}^N [p_i^{(k)}]^2,$$
and $\Pi$ is the purity $\mathrm{Tr}(\rho^2)$ of the state $\rho$.
Clearly, $\frac{1}{N} \leq \pi_k \leq 1$, and similarly for $\Pi$.  Then subtracting the excess we can see that
\begin{equation}
\label{eq:pisum}
\sum_{k=1}^M \pi_k \leq \Pi + 1 - \frac{N + 1 - M}{N} \leq \frac{N - 1 + M}{N}.
\end{equation}

Since the product of a set of positive numbers whose sum is constrained is maximized when the numbers are all equal,
we can use (\ref{eq:pisum}) to write
\begin{equation}
\label{eq:piprod}
\prod_{k=1}^M \pi_k \leq \left( \frac{N-1+M}{NM} \right)^M.
\end{equation}
Because $H_k \geq - \ln \pi_k$, \cite{maassen} this gives us the relation
\begin{equation}
\label{eq:result}
\sum_{k=1}^M H_k \geq M \log \frac{NM}{N-1+M}.
\end{equation}

\begin{figure}[t]
\centering
\includegraphics{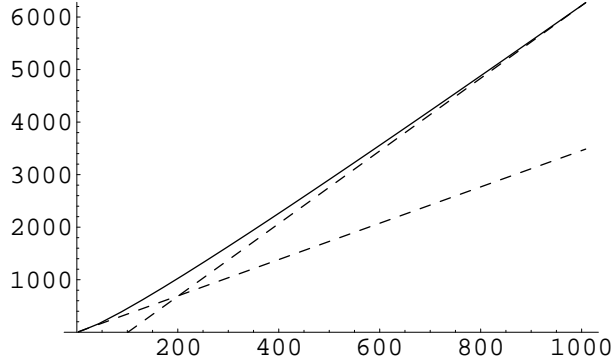}
\caption{Equation (\ref{eq:result2}) vs. $M$ for $N=1009$.  The dashed lines represent the weaker lower bounds
(\ref{eq:weak1}) and (\ref{eq:weak2}).}
\label{fg:graph}
\end{figure}

S\'{a}nchez \cite{sanchez2} showed that his bound (\ref{eq:san}) for complete sets, where the dimension $N$ is even,
cannot be tight as the bound $H_k \geq - \log \pi_k$ can only be tight for $\pi_k = 1/m,~m \in \mathbb{N}$.  For
$\frac{1}{m} \leq \pi_k \leq \frac{1}{m-1},~ z \in \mathbb{N}$ he derived \cite{sanchez3} the convex inequality
$$ H_k \geq \log m - (m-1)(m \pi_k - 1) \log \frac{m}{m-1} $$
This can be applied to strengthen relationship (\ref{eq:result}), yielding
\begin{equation}
\label{eq:result2}
\sum_{k=1}^M H_k \geq M \left[ \log m - (m-1)\left(m \frac{N+M-1}{NM} - 1 \right)\log \frac{m}{m-1} \right]
\end{equation}
where
$$m = \mathrm{ceil}\left(\frac{NM}{N+M-1}\right).$$
Figure \ref{fg:graph} plots this bound with respect to $M$.

Relations (\ref{eq:result}) and (\ref{eq:result2}), for
$M \sim N / \log N$ (near the intersection of (\ref{eq:weak1}) and (\ref{eq:weak2})) provide a significantly
improved bounds.  The behavior at the end points $M=1$ and $M=N+1$ is correct as well.  Relation (\ref{eq:result})
is only marginally weaker, and then only for $NM/(N+M-1) \not\in \mathbb{N}$, and has a somewhat more convenient
form.

\bibliographystyle{unsrt}
\bibliography{entropicUncertainty}

\section*{Acknowledgements}
I would like to thank my mentors Drs. John Preskill and Patrick Hayden for their direct support in this
project.  I would also thank Rohit Thomas and Peter Knophf for their help with untangling some nasty math,
and the donor for the Victor Neher named SURF fellowship for financial support.

\end{document}